\documentclass[a4paper,11pt]{article}
\usepackage[a4paper, total={17cm, 25cm}]{geometry}
\usepackage{fancyhdr}
\usepackage{xcolor}
\usepackage{amsmath}
\fancypagestyle{plain}{
\fancyhf{}
\fancyhead[RH]{\thepage}
\fancyfoot[CF]{28th Texas Symposium on Relativistic Astrophysics \\ Geneva, Switzerland -- December 13-18, 2015}

}
\begin{document}

\title{\bf Comments on the Effect of Frame Dragging}

\author{\.Ismail \"Ozbak{\i}r$^1$  and Kadri Yakut$^1$  \\
$^1$ University of Ege, Faculty of Science, Department of Astronomy and Space Science, \.Izmir, Turkey}

\date{February 15, 2016}

\maketitle

\begin{abstract}
The effect of the frame dragging on the equation of motions, depends on the approaches that have been considered. Accordingly, additional force terms may appear or disappear. To understand the effect of radial and non-radial perturbations that may exist in the case of mass distribution of spherical shell form, modifications should be predicted to the approaches of the components of elastic tensor $E^{\mu\nu}$ given in Bass and Pirani\cite{Bass1955}. Different relations between $\omega$ and $\omega'$ have been discussed.
\end{abstract}


\section{Introduction}

Einstein's general theory of relativity (GR) predicts many astronomical phenomena (Einstein, 1916a, 1916b, 1918).
Frame dragging  effect is one of the prediction of GR and first studied by J. Lense  and  H. Thirring \cite{LT1918}\cite{Thirring1918}\cite{Thirring1921}
This effect is mainly deform and drag the spacetime around rotating massive objects.
Gravity  Probe  B  gyroscopes have detected frame  dragging  effect around the Earth \cite{Everitt2011}
By using a linear approach method Hans Thirring \cite{Thirring1918}\cite{Thirring1921} gives the effect of spherical mass distribution with a radius $\textit{a}$ that rotates with a constant $\omega$ speed around an axis, on the equation of motions as

\begin{equation}
g_{\mu\nu}=-\delta_{\mu\nu}+\gamma'_{\mu\nu}-\frac{1}{2}\delta_{\mu\nu}\gamma'_{\alpha\alpha}
\end{equation}

\begin{equation}
\gamma'_{\mu\nu}=-\frac{\chi}{4\pi}\int\frac{T_{\mu\nu}(x,y,z,t-r)}{R}dV_0
\end{equation}
where $\chi=8\pi k$; k is the gravity constant and $dV_0$ is the spatial volume element. There is an inconsistency, on the other hand, between Thirring's assumptions because of the violation of energy -– momentum conservation law:

\begin{equation}
T^{\mu\nu}_{;\nu}=0
\end{equation}

Bass and Pirani \cite{Bass1955} add the elastic term $E^{\mu\nu}$ to the energy--momentum tensor to solve this inconsistency.

\begin{equation}
T^{\mu\nu}=\rho v^{\mu}v^{\nu}+E^{\mu\nu}
\end{equation}

In their calculation they used the approach that Thirring applied on the metric tensor components. In addition, they define the mass density of the spherical shell in the form of $\rho(\theta)=\rho_0(1+Na^2\omega^2\sin^2 \theta)$ which depends on the latitude. When the constant term \textit{N} assumed to be \textit{N=−1}, then this corresponds to a mass distribution that compansates for the spherial relativistic mass increase and represents a uniform mass distribution in the reference frame where the shell is rotating.

\section{A new additional term in the solution}

In the work of Bass and Pirani \cite{Bass1955} the approaches on the elastic energy--momentum terms $E^{\mu\nu}$ can be modified. Energy-momentum conservation equation, $T^{\mu\nu}_{;\nu}=0$, estimations examiations reveal that only the $E^{\mu\nu}$ components lattitude depended partial derivation have been taken into consideration. This means that the frame dragging effects, that may arise due to radial and non--radial mass fluctuations on the spherical shell have been neglected.
In addition, the estimation of the $\gamma'_{44}$ and $\gamma'_{24}$ perturbation terms in the metric components, the value of $(\frac{dx^4}{ds})^2$  according to the $\omega$ term for different series approaches cause variations in the equation of motion. When $\gamma'_{44}$ and $\gamma'_{24}$ terms are estimated with second order (\textit{S2}) and first order (\textit{S1}) approaches (which is Thirring's approach) than the equation of motions are found as

\begin{subequations}
	\begin{gather}
	\ddot{x}=-\frac{8kM}{3a}\omega\dot{y}+\frac{2kM(1+N)}{15a}\omega^2x\\
\ddot{y}=\frac{8kM}{3a}\omega\dot{x}+\frac{2kM(1+N)}{15a}\omega^2y\\
\ddot{z}=\frac{4kM(1+N)}{15a}\omega^2z
	\end{gather}
\end{subequations}

If \textit{S1} and \textit{S2} approaches are applied to the $\gamma'_{44}$ and $\gamma'_{24}$ terms; respectively, then

\begin{subequations}
	\begin{gather}
\ddot{x}=-\frac{8kM}{3a}\omega\dot{y}\\
\ddot{y}=\frac{8kM}{3a}\omega\dot{x}\\
\ddot{z}=0
	\end{gather}
\end{subequations}

If on the other hand, \textit{S1} and \textit{S2} approaches are applied to the $\gamma'_{44}$ and $\gamma'_{24}$ terms then one can get the equations

\begin{subequations}
	\begin{gather}
\ddot{x}=-\frac{8kM}{3a}\omega\dot{y}\ {\color{red}{-\frac{4kMa\omega^2(7+2N)}{45}\omega\dot{y}}}+\frac{2kM(1+N)}{15a}\omega^2x \\
	\ddot{y}=-\frac{8kM}{3a}\omega\dot{x}\ {\color{red}{-\frac{4kMa\omega^2(7+2N)}{45}\omega\dot{x}}}+\frac{2kM(1+N)}{15a}\omega^2y \\
\ddot{z}=-\frac{4kM(1+N)}{15a}\omega^2z
	\end{gather}
\end{subequations}

The relation of $\omega'$ angular velocity of a particle with the $\omega$ that has been presented in Thirring \cite{Thirring1918} has not ben considered in the work of Bass and Pirani \cite{Bass1955}. We can get this relation without changing any conditions that have been applied to the elastic tensor $E^{\mu\nu}$ as

\begin{multline}
\ddot{x}=2\left[ \omega'\left(1+\frac{2kM}{a} \right)-\omega\frac{4kM}{3a}+\frac{2kM}{3}a\omega^2\omega'\left( 1+\frac{2}{5a^2}(1+N)(x^2+y^2-z^2) \right) \right]\dot{y}\\
+\left\{ \omega'^2\left(1+\frac{2kM}{a} \right)-\omega\omega'\frac{8kM}{3a}+\omega^2\frac{2kM(1+N)}{15a}+\frac{2kM}{3}a\omega^2\omega'\left( 1+\frac{2}{5a^2}(1+N)(x^2+y^2-z^2) \right) \right\}x \\
-\frac{8kM(1+N)}{15a}y\omega^2z\omega'\dot{z}
\end{multline}

\begin{multline}
	\ddot{y}=-2\left[ \omega'\left(1+\frac{2kM}{a} \right)-\omega\frac{4kM}{3a}+\frac{2kM}{3}a\omega^2\omega'\left( 1+\frac{2}{5a^2}(1+N)(x^2+y^2-z^2) \right) \right]\dot{x}\\
	+\left\{ \omega'^2\left(1+\frac{2kM}{a} \right)-\omega\omega'\frac{8kM}{3a}+\omega^2\frac{2kM(1+N)}{15a}+\frac{2kM}{3}a\omega^2\omega'\left( 1+\frac{2}{5a^2}(1+N)(x^2+y^2-z^2) \right) \right\}y\\
	+\frac{8kM(1+N)}{15a}x\omega^2z\omega'\dot{z}
\end{multline}

\begin{multline}
\ddot{ z }=-\frac{4kM(1+N)}{15a}\omega^2z-\frac{8kM(1+N)}{15a}\omega^2\omega'(x\dot{ y }-\dot{ x }y)z-\frac{4kM(1+N)}{15a}(x^2+y^2)\omega^2\omega'^2z
\end{multline}

When we apply the approach of $\omega^2\omega'\sim0$ in Thirring \cite{Thirring1918} work to the above equations (cf. \cite{Thirring1921} eq. (25)) we obtain the equation of motions that depends on the mass density term N as

\begin{alignat}{2}
\ddot{x}=2\left[ \omega'\left(1+\frac{2kM}{a} \right)-\omega\frac{4kM}{3a} \right]\dot{ y }+\left\{  \omega'^2\left(1+\frac{2kM}{a} \right)-\omega\omega'\frac{8kM}{3a}+\omega^2\frac{2kM(1+N)}{15a} \right\}x\\
	\ddot{y}=-2\left[ \omega'\left(1+\frac{2kM}{a} \right)-\omega\frac{4kM}{3a} \right]\dot{ x }+\left\{  \omega'^2\left(1+\frac{2kM}{a} \right)-\omega\omega'\frac{8kM}{3a}+\omega^2\frac{2kM(1+N)}{15a} \right\}y\\
	\ddot{z}=-\frac{4kM(1+N)}{15a}\omega^2z
\end{alignat}

Numerical solutions can be applied to these equations for the condition \textit{N=−1} most interesting case that appears in \cite{Bass1955}.

\paragraph{}

\section{Results}
In this work, we study the partial derivation of the elastic tensor $E^{\mu\nu}$ that appears in the energy--momentum conservation equations, $T^{\mu\nu}_{;\nu}=0$, and in other relations that depends on it. We also aim to study other coordinate derivatives in addition to the latitude dependence that has been studied in \cite{Bass1955}. Therefore, radial and non--radial perturbation fluctuations that may occur in the shell mass distribution will be studied. Moreover, we will work on the posible solutions of equation of motion by taking different approaches into account. Therefore, the relation of space--time frame dragging with different types of pulsations will be analysed in details.
In the work of \cite{Bass1955}, $\omega$ and $\omega'$ dragging coefficient relation presented in \cite{Thirring1918} has been investigated. In this study, different relations between  $\omega$ and $\omega'$ have been discussed. In our future work, we plan to study the effect of frame dragging on the neutron star interiors, binary stars with a neutron star and on the gravitational waves which is recently detected by LIGO \cite{LIGO2016a}.

\paragraph{}
\textbf{Acknowledgments}\\
This study was supported by the Turkish Scientific and Research Council (T\"UB\.ITAK 113F097).
The current study is a part of PhD thesis by \.I \"Ozbak{\i}r.

\paragraph{}

\end{document}